\documentclass[12pt]{iopart}
\usepackage{iopams}
\usepackage{indentfirst}
\usepackage{braket}
\usepackage{makeidx}
\usepackage{slashed}
\usepackage{graphicx}
\usepackage{xcolor}
\usepackage{hyperref}
\usepackage[normalem]{ulem}

\newcommand{\ke}[1]{\vert{#1}\rangle}
\newcommand{\brat}[1]{\langle{#1}\vert} 
\newcommand{\op}[2]{\ke{#1}\!\brat{#2}}
\newcommand{\cH}{{\cal H}}
\newcommand{\cB}{{\cal B}}
\newcommand{\cL}{{\cal L}}
\begin{document}
\title{Magnetic field induced symmetry breaking in nonequilibrium quantum networks}
\author{Juzar Thingna$^{1}$, Daniel Manzano$^{2}$ and Jianshu Cao$^{3}$}
\address{%
$^1$Center for Theoretical Physics of Complex Systems, Institute for Basic Science (IBS), Daejeon 34126, Republic of Korea}
\address{%
$^2$Universidad de Granada, Departamento de Electromagnetismo y F\'isica de la Materia and Instituto Carlos I de F\'isica Te\'orica y Computacional, Granada 18071, Spain}
\address{%
$^3$Massachusetts Institute of Technology, Chemistry Department. Cambridge, Massachusetts 02139, USA}
\ead{\mailto{jythingna@ibs.re.kr}; \mailto{manzano@onsager.ugr.es}; \mailto{jianshu@mit.edu}}

\begin{abstract}
We study the effect of an applied magnetic field on the nonequilibrium transport properties of a general cubic quantum network described by a tight-binding Hamiltonian with specially designed couplings to the leads that preserve open-system symmetries. We demonstrate that the symmetry of open systems can be manipulated by the direction of the magnetic field. Starting with all the symmetries preserved in absence of a field, the anisotropic and isotropic fields systematically break the symmetries, influencing all nonequilibrium properties. For simple cubic systems, we are able to identify the steady states that comprise of pure states, bath-dependent states (nonequilibrium steady states), and also nonphysical states. As an application, we show numerically for large cubic networks that the symmetry breaking can control nonequilibrium currents and that different environmental interactions can lead to novel features which can be engineered in artificial super-lattices and cold atoms.
\end{abstract}
\maketitle
\section{Introduction}
\label{sec:Intro}
Symmetries in closed quantum systems have been a cornerstone in modern physics, introducing constraints to simplify and solve complex many-body systems. In recent years, inspired by their predictive capabilities, the basic principles of symmetries have been extended to \emph{open} quantum systems. It has been proved that the dynamics of an open system can present several fixed points (asymptotic states) if the system is invariant under a symmetry \cite{BucaNJP12} operation. This degeneracy of the dynamics generator determines the thermodynamic properties of the system and can potentially give rise to dynamical phase transitions \cite{ManzanoPRB14, ManzanoAdvPhys18}. The symmetry related phenomena are surprisingly robust under weak symmetry breaking and are reflected in the metastable states \cite{ThingnaSciRep16, MacieszczakPRL16} instead of the asymptotic states.

Despite these novelties \cite{AlbertPRA14}, little is known about how the breaking of symmetries affects the nonequilibrium properties of the system. In order to address this issue, we focus on cubic networks placed under a temperature bias, driving the system far from equilibrium. Cubic networks are one of the simplest closed systems with symmetries, found in many natural materials \cite{Ashcroft76}. Recently, several experimental setups have been made accessible to study quantum transport in \emph{square lattices}. Examples are optical lattices, where several transport experiments have been already performed \cite{SchneiderNatPhys12, HildPRL14}, and  two-dimension ion traps with tunable spin-spin interactions \cite{KumphNJP11, BrittonNat12, WilsonNat14}. Theoretically, due to the presence of open-system symmetries in square lattices both ballistic and diffusive subspaces can coexist \cite{ZnidaricPRL13, ManzanoNJP15}.

Inspired by these results and the simplicity of cubic networks, we aim to address if the closed-system symmetry can help us simplify the open-system nonequilibrium problem. We analyze the cubic networks driven far from equilibrium and obtain the underlying open-system symmetry operators and consequently the steady-state reduced density matrices that reflect the symmetries. Furthermore, we break the closed-system symmetry using Abelian magnetic fluxes \cite{GoldmanRPP14} and show how this affects the open system leading to a control of nonequilibrium transport properties. In particular, we find that the direction of the magnetic field, and not its magnitude, plays a crucial role to control transport properties.

The rest of the paper is arranged as follows: In Sec.~\ref{sec:model} we describe the basic model of an open quantum cubic network in presence of a magnetic field. Section~\ref{sec:symm} reviews the basic notions of open-system symmetries and the non-triviality of obtaining the symmetry operators for open systems. We then treat analytically the simplest possible case of a system arranged as a square or cube in Sec.~\ref{sec:analytics}. Here, due to the analytically obtainable symmetry operators, we get the initial condition independent steady states. We then in Sec.~\ref{sec:numerics} extend our model to quasi-one dimensional cubic networks and explore the transport numerically for different system-bath connections. In both the analytics and numerics we find that the magnetic field direction breaks symmetries that can dramatically affect the transport properties of the system. Finally, in Sec.~\ref{sec:conclusion} we conclude.

\section{Cubic Network in a Magnetic Field}
\label{sec:model}
\begin{figure}
\centering
	\includegraphics[width=\columnwidth]{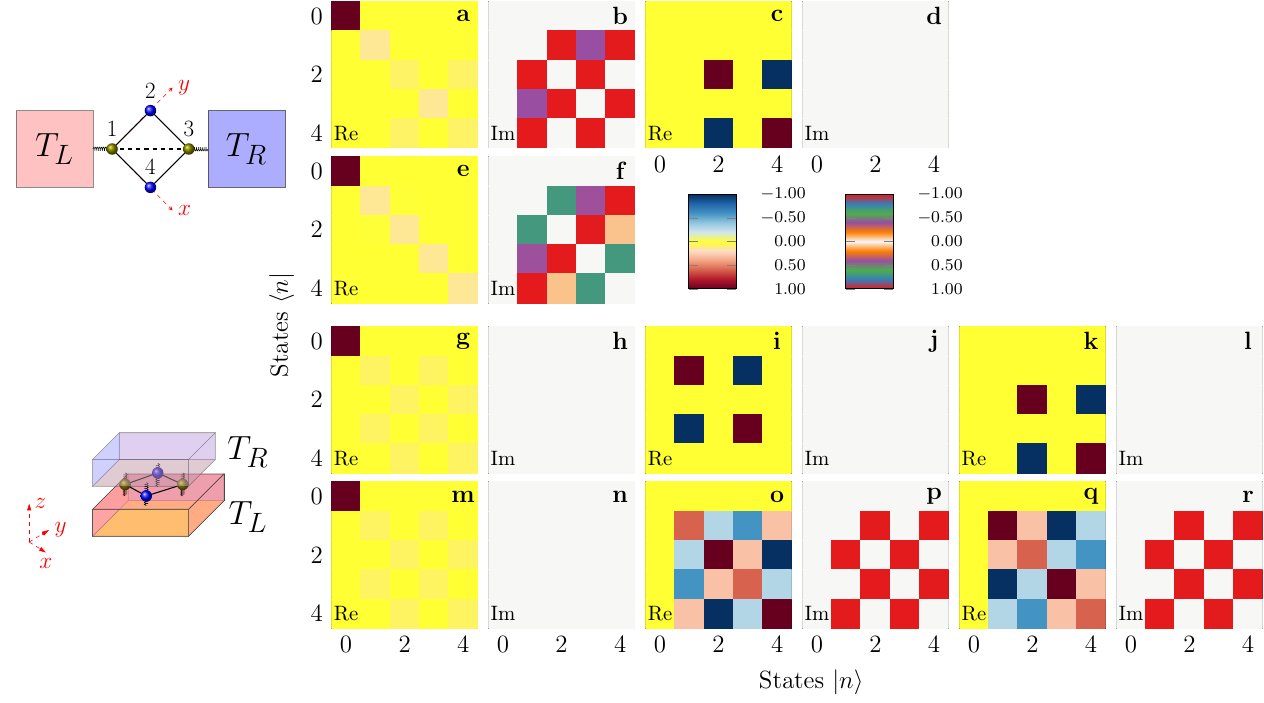}
\caption{(Color online) Various steady states for a single plaquette (square) system. The edge coupled plaquette (top view) is illustrated in the first column top and, the face coupled plaquette (side view) is shown in the first column bottom. Panels a-d correspond to edge coupled plaquette without a magnetic field, whereas panels e-f have an anisotropic ($z$ direction) magnetic field. Panels g-r correspond to the face coupled plaquette without (panels g-l) and with (panels m-r) an anisotropic magnetic field. Each panel is either the real or imaginary part of the RDM (marked bottom left corner). Each panel is normalized by the largest value and is the RDM plotted in the single particle site basis $i$ ($i=1,\cdots,4$) with $0$ being the vacuum state of the system. The master equation in this representation can be simplified, see~\ref{sec:single-exc}. The parameters used for the calculations are: $\varepsilon_{1}=\varepsilon_{3}=1$, $\varepsilon_{2}=\varepsilon_{4}=1.2$, $t_1=t_2=t_3=1$, $B=\pi/4$, $\Gamma=0.1$, $\epsilon_0=1$, $T_L=0.6$, and $T_R=0.4$.}
\label{fig:analy}
\end{figure}

\begin{figure}[t!]
\centering
	\includegraphics[width=0.8\columnwidth]{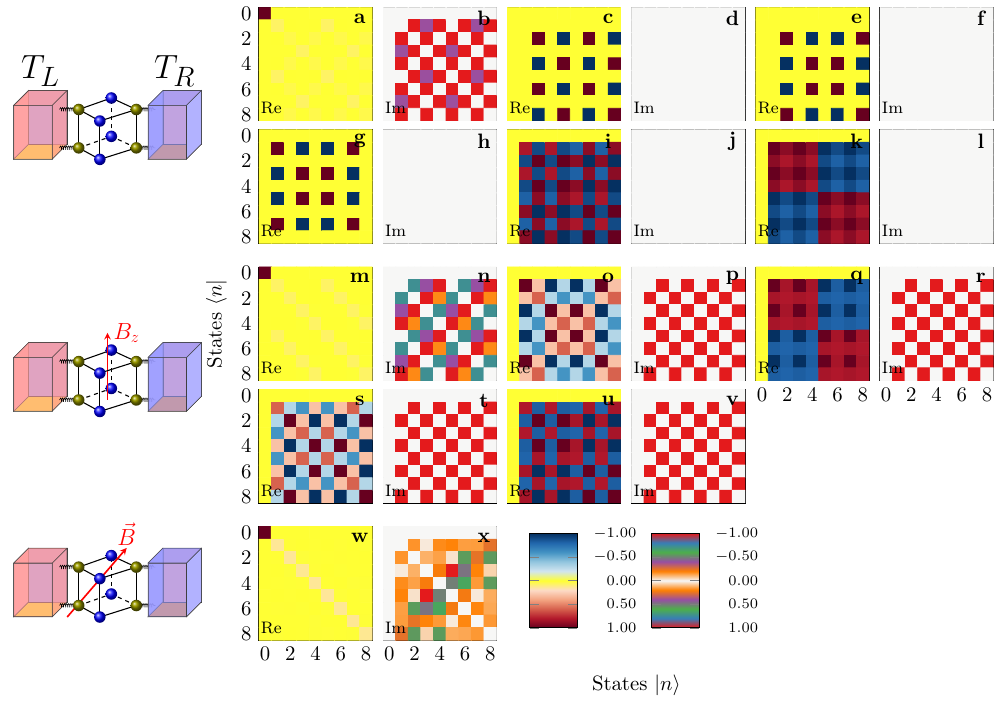}
\caption{Various steady states for two plaquette (cube) system edge connected to the reservoirs. Panels a-l are the RDMs (real and imaginary parts) for the system (see top illustration) in the absence of magnetic field. Panels m-v are in the presence of an anisotropic field (middle illustration) and panels w-x are in presence of an isotropic field (bottom illustration). Each panel is normalized by the largest value. The parameters used for the calculations are: $\varepsilon_{1,5}=\varepsilon_{3,7}=1$, $\varepsilon_{2,6}=\varepsilon_{4,8}=1.2$, $t_1=t_2=t_3=1$, $B=\pi/4$, $\Gamma=0.1$, $\epsilon_0=1$, $T_L=0.6$, and $T_R=0.4$.}
\label{fig:2RL}
\end{figure}

\begin{figure}[bt!]
\centering
	\includegraphics[width=0.8\columnwidth]{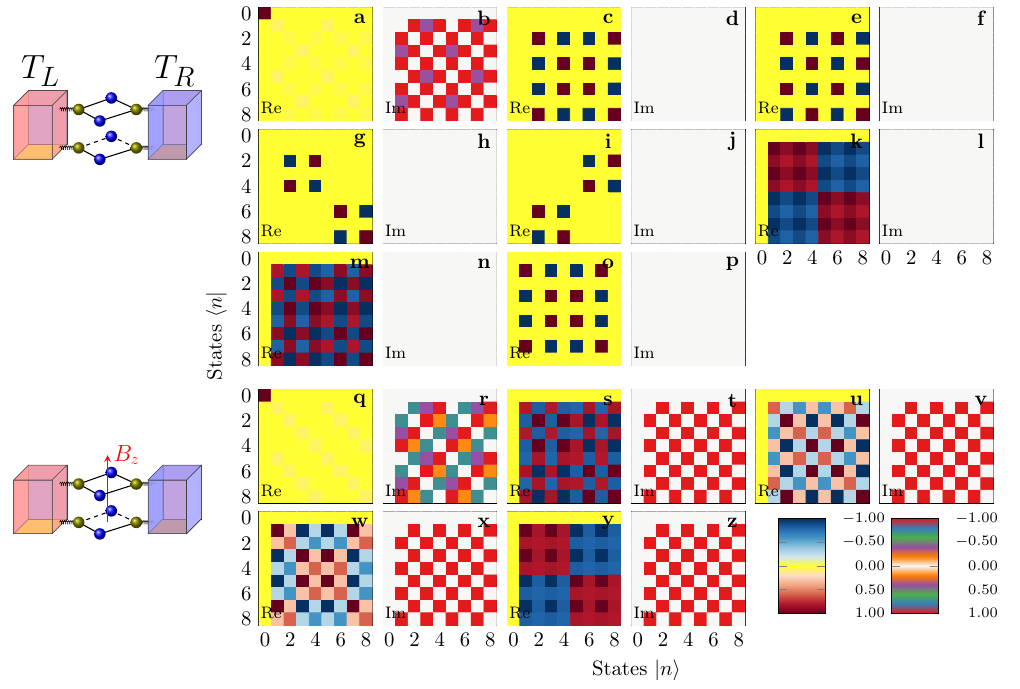}
\caption{Various steady states for two plaquette (cube) system without vertical links (stacked) edge connected to the reservoirs. Panels a-p are the RDMs (real and imaginary parts) for the system (see top illustration) in the absence of magnetic field. Panels q-z are in the presence of an anisotropic/isotropic field (see bottom illustration). Each panel is normalized by the largest value. All parameters are the same as the ones used in Fig.~\ref{fig:2RL} except $t_3=0$.}
\label{fig:2RS}
\end{figure}

The system of interest is a cubic network of non-interacting fermions subjected to an external magnetic field illustrated in Figs.~\ref{fig:analy}--~\ref{fig:2RG},~\ref{fig:2RS}, and \ref{fig:3D}. The tight-binding representation of the system Hamiltonian is given by,
\begin{equation}
\label{eq:1}
H=-\sum_{r} \varepsilon_r c^{\dagger}_r c_r -\sum_{r, d =1,2,3} t_d U^d(\hat{r})c^{\dagger}_{r+d}c_r + \mathrm{h.c.},
\end{equation}
where $r=(x, y, z)$ is the position of the lattice sites, $\varepsilon_r$ is the onsite energy at position $r$, and $t_{d}$ is the tunneling strength in the direction $d$ with $d=1, 2,$ and $3$ corresponding to $X, Y,$ and $Z$ direction respectively. $c^{\dagger}$ and $c$ are the fermionic creation and annhilation operators, respectively. The hopping element is isotropic. The onsite energies are different for the neighboring sites in the $X-Y$ plane and homogeneous in the $Z$ direction. This leads to the lattice having a period $2a$ ($a$ being the norm of the primitive vector) in the $X-Y$ plane causing the isotropic magnetic field to effect the couplings in all directions. The magnetic field effects are absorbed in the tunneling phases $U^1(r) = 1$, $U^2(r) = \exp \left[ i2\pi B\left(x-y-\frac{1}{2}\right) \right] $, and $U^3(r)= \exp \left[ -i\pi B\left(x-y\right)\right]$ with the cube length $a=1$  \cite{note1}. The phases are obtained under the Hasegawa gauge \cite{HasegawaJPSJ90,HasegawaPhysicaC91,LinPRB96,BurrelloJPA17,Burrelloreview17} that can be experimentally realized in artificial super-lattices \cite{GeislerPRL04,MelintePRL04,DeanNat13,PonomarenkoNat13} or cold atoms \cite{GoldmanRPP14,AidelsburgerPRL13,MiyakePRL13}.

Our system is connected to two reservoirs as shown in the illustrations of Figs.~\ref{fig:analy}--~\ref{fig:2RG},~\ref{fig:2RS}, and \ref{fig:3D} allowing us to explore the effects of magnetic fields on spatially symmetric nonequilibrium dissipative systems. We engineer the system-bath coupling Hamiltonian
\begin{equation}
H_{S\alpha} = \sum_{l} c^{\dagger}_l \Pi_{i\neq l} c_ic_i^{\dagger} \sum_{k}\nu_k d_{k,\alpha}+\mathrm{h.c.}
\end{equation}
with $l$ index characterizing the sites connected to the reservoir $\alpha = L, R$ ($L=$ left reservoir, $R=$ right reservoir). It has been observed that engineering of baths and their coupling to ultracold atoms is possible and leads to interesting new features not observed in natural systems \cite{DiehlNatPhys08, MullerAAMOP12}. The fermionic operators $d$, $d^{\dagger}$ correspond to the annihilation and creation operators of the reservoir, $H_{\alpha} = \sum_{k} \epsilon_k d_{k,\alpha}^{\dagger}d_{k,\alpha}$, with $\nu_k$ being the system reservoir coupling strength. The specific form of the coupling allows the reservoirs to locally inject or extract a particle between the single particle states and the ground state. It is important to note here that for the extensively studied scenario of electron hopping between the system and reservoir $H_{S\alpha} = \sum_{l}c^{\dagger}_l \sum_k\nu_k d_{k,\alpha}+\mathrm{h.c.}$. This coupling Hamiltonian does not have any open-system symmetries and gives only one \emph{unique} steady state. This is perhaps one of the crucial reasons that even for a simple cubic structure the effect of the magnetic field \emph{direction} has not been observed in the literature.

The dynamics of the reduced density matrix (RDM) is governed by a Markovian dissipative Lindblad quantum master equation \cite{Alicki2007, Breuer2007, Gorini1976, Lindblad1976} that reads,
\begin{equation}
\label{eq:2}
\frac{d\rho}{dt} = \cL\rho= - i [H,\rho] + \sum_{k=1,2 \atop \alpha={ L, R}} \Gamma_{\alpha k} \mathcal{D}[A_{\alpha k}]\rho.
\end{equation}
Where $\cL$ is the Liouvillian superoperator of the system, $\mathcal{D}[A_{\alpha k}]\rho = A_{\alpha k}^{\phantom{\dagger}}\rho A_{\alpha k}^{\dagger}-\frac{1}{2}\{A_{\alpha k}^{\dagger}A_{\alpha k}^{\phantom{\dagger}},\rho\}$,
the Lindblad operators $A_{\alpha k}$ ($A_{\alpha 1}=\sum_{n} c_n^\dagger \Pi_{i\neq n} c_ic_i^{\dagger}\equiv  A_{\alpha 2}^\dagger$) allow local injection ($k=1$) or extraction ($k=2$) of particles into the system. The sum over $n$ in the Lindblad operators is restricted to $n=1,5,9,\cdots$ ($n=1,\cdots,4$) for $\alpha = L$ and $n=3,7,11,\cdots$ ($n=N-3,\cdots,N$) for $\alpha = R$ for edge (face) connected baths. The injection rate $\Gamma_{\alpha 1} = \Gamma f_{\alpha}(\epsilon_0)$ with $f_{\alpha}(\epsilon_0) = \left(\exp[\beta_\alpha \epsilon_0]+1\right)^{-1}$ being the Fermi-Dirac distribution. The ratio of the rates $\Gamma_{\alpha 1}/\Gamma_{\alpha 2} = \exp [-\beta_{\alpha}\epsilon_{0}]$ obeys local-detailed balance where $\beta_{\alpha}=1/k_{{\rm B}}T_{\alpha}$ represents the inverse temperature of the $\alpha$-th bath and $\epsilon_0$ is the characteristic energy of the bath. The system-reservoir coupling ensures that without the magnetic field the generator can be split into invariant subspaces that lead to multiple steady-states \cite{BucaNJP12, ManzanoAdvPhys18, ThingnaSciRep16, AlbertPRA14}.


\section{Symmetry Analysis}
\label{sec:symm}
In this section we briefly review the relation between symmetry and invariant subspaces of the system dynamics (a more detailed explanation can be found in Refs. \cite{BucaNJP12, ManzanoAdvPhys18}). For a specific set of generators of the Markovian dynamics of our system $\left\{H, \left\{ A_{\alpha,k} \right\}_{k=1,2 \atop \alpha={ L, R}} \right\}$, if there is a unitary operator $U$ s.t. 

\begin{equation}
\label{eq:symmcomm}
\left[ U,H\right] = \left[ U, A_{\alpha,k}\right]=0 \qquad\forall (\alpha,k),
\end{equation}
then the system has a {\it strong symmetry}. In this case, we can prove that there are several invariant subspaces of the system dynamics, and therefore several orthogonal steady-states. To do so, we decompose the Hilbert space of the pure system, $\cH$, by the spectral decomposition of $U$. If our symmetry operator has $n_U$ eigenvalues we have that $U\ket{\psi_i^{(k)}}=\exp{(i\theta_i)} \ket{\psi_i^{(k)}}$ ($i=1,\dots, n_U$) and $k=1,\dots,d_i$ with $d_i$ being the dimension of the subspace corresponding to the eigenvalue $\exp{(i\theta_i)}$. Using this we decompose the Hilbert space in the form $\cH=\bigoplus_i \cH_i$, with  $\cH_i= $span$ \left\{ \ket{\psi_i^{(k)}}: {k=1, \dots, d_i} \right\}$. The operator space can be decomposed in a similar way as $\cB(\cH)=\bigoplus_{ij} B_{ij}$, being $B_{ij}=$ span $\left\{ \op{\psi_i^{(k)}}{\psi_j^{(l)}}: k=1,\dots, d_i; l=1,\dots,d_j \right\} $. 

We define now the left and right symmetry superoperators acting on the operators space as $U_L O=U O$ and $U_RO=O \,U$, for any $O\in \cB(\cH)$. It is clear that the subspaces  $B_{ij}$ are the eigenspaces of these superoperators, meaning that  $U_L B_{ij} \subseteq B_{ij}$ as well as $U_R B_{ij} \subseteq B_{ij}$. As the system dynamics is generated by a superoperator $\cL$ in the form of Eq. \ref{eq:2}, and all the dynamics generators commute with the symmetry operator $U$ it follows that the subspaces $B_{ij}$ are invariant under the system dynamics $\cL B_{ij}\subseteq B_{ij}$.

There is also the possibility of  a superoperator $S$ that commutes with the system Liouvilian $\left[ S,\cL \right]=0$ without having a strong symmetry associated. This is called a {\it weak symmetry} and we can show, by the same reasoning as with the strong symmetries, that a weak symmetry also implies the existence of invariant subspaces.  

As we are working only with bounded systems we can apply Evans' Theorem \cite{Evans} to prove that in each subspace in the form $B_{ii}$ there is at least one fixed point of the dynamics (eigenvectors of $\cL$ with zero eigenvalue). It has also been proposed in Ref \cite{BucaNJP12} the possibility of finding fixed points in the subspaces $\cB_{ij}$ with $i\neq j$. These subspaces do not contain physical states as they include only zero trace operators. In this work we provide the first example of these non-physical fixed points. 

The problem of finding the symmetry decomposition of a given system is still open. Thus, finding all the symmetry operators and invariant subspaces of an arbitrary system is a challenging task. In other words, without knowledge of all symmetry operators it is impossible to find all the invariant subspaces and consequently finding the RDMs within these subspaces is highly non trivial. Moreover, any naive attempt at numerically computing the steady-state RDMs will result in a matrix that is a linear combination of the \emph{true} RDMs, and is generally not a true density matrix (see~\ref{sec:degenerated} for more details on how to overcome this problem).

\section{Analytically Tractable Systems}
\label{sec:analytics}
Given the cubic network in a magnetic field, we first provide the steady state solutions ($\rho_{ss} = \lim_{t\rightarrow \infty}\rho$) utilizing the underlying symmetries of the system. In the cases discussed here, using the symmetry operators we were able to obtain all the far from equilibrium steady states (states that are independent of initial condition), which is nontrivial to obtain in degenerated open quantum systems (See~\ref{sec:degenerated}).

Figure~\ref{fig:analy} shows the various steady states for a single plaquette (square) illustrated in the first column. The edge coupled (top) and face coupled (bottom) reservoirs posess different open-system symmetries, giving rise to different steady states. The closed-system possess 4 symmetries (rotation by $0$, $\pi$ and flip about both diagonals) given by the plane symmetries of a square with different adjacent vertices. Not all of these four closed-system symmetries translate to open-system symmetries and the coupling of the system to the bath plays an important role [see Eq.~\ref{eq:symmcomm}]. In case of edge coupled plaquette, the open-system symmetry that survives in absence of magnetic field is the flip symmetry along the diagonal (the axis of symmetry is indicated by the dashed line in the illustration). The corresponding unitary symmetry operator is $U=\exp\left(\op{2}{4} + \op{4}{2} \right)$. The symmetry operator block diagonalizes the Liouvillian giving rise to two steady states as indicated in panels a-d, obtained analytically and represented in the single particle basis. Out of the two states, one is a nonequilibrium steady state (NESS) that depends on the properties of the reservoir (panels a-b) and the other one is a pure state independent of the reservoir properties  (panels c-d). The pure state turns out to be anti-symmetric under the exchange of states $|2\rangle$ and $|4\rangle$ as seen in panel c. Such states carry zero current \cite{EngelhardtPRB19} and are also referred to as dark or compact localized states \cite{MaimaitiPRB17}. In presence of an anisotropic ($z$-direction) or isotropic magnetic field (panels e-f) the edge coupled system displays a single NESS that differs from the one obtained in the absence of magnetic field (panels a-b). 

The face coupled single plaquette in the absence of magnetic field (panels g-l) gives three physical steady states with two pure states that are anti-symmetric with respect to exchange of sites $|1\rangle - |3\rangle$ (panel i) and $|2\rangle - |4\rangle$ (panel k). The corresponding symmetry operators are $U_1=\exp \left( \op{2}{4}+\op{4}{2} \right) $ and $U_2=\exp \left( \op{1}{3}+\op{3}{1} \right) $. In the absence of a magnetic field all steady states are real. Once the magnetic field (anisotropic or isotropic) is applied we break the flip symmetry along the diagonal but the $\pi$ rotation symmetry still persists. Additionally, a new non-topological symmetry  emerges that gives us three steady state solutions even in the presence of a magnetic field (panels m-r). In this case all the states are physical and we still find only one NESS. The two pure states (panels o-r) are non trivial, i.e., either symmetric or anti-symmetric w.r.t the sites. Interestingly in this case the NESS is invariant under the influence of a magnetic field (in stark contrast to the edge coupled case).

\begin{figure}[t]
\centering
	\includegraphics[width=0.8\columnwidth]{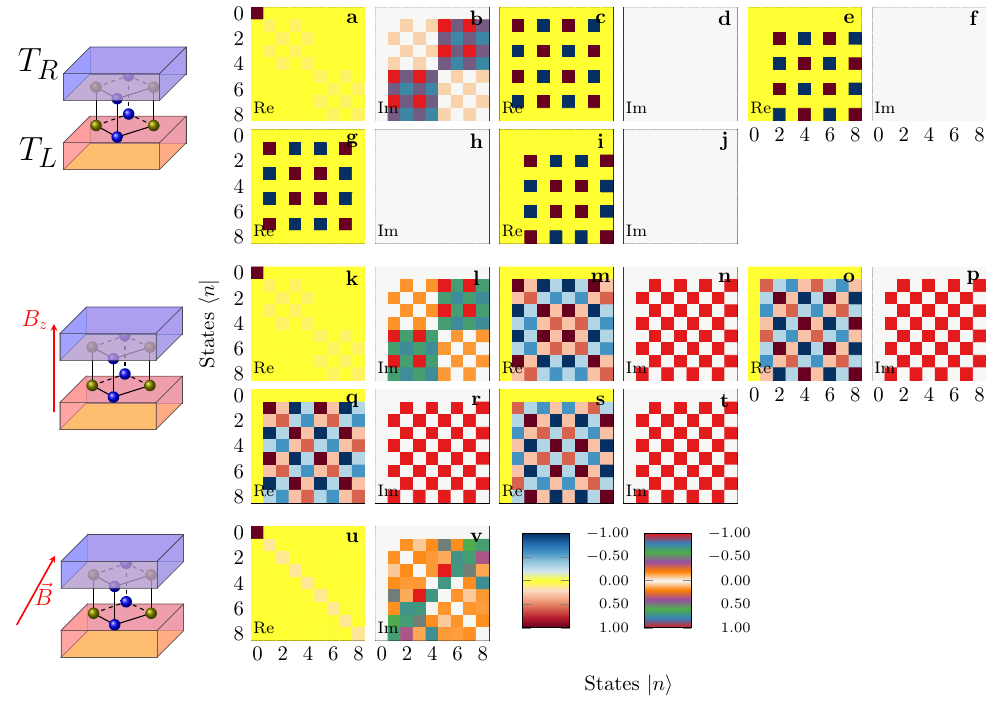}
\caption{Various steady states for two plaquette (cube) system face connected to the reservoirs. Panels a-j are the RDMs (real and imaginary parts) for the system (see top illustration) in the absence of magnetic field. Panels k-t are in the presence of an anisotropic field (middle illustration) and panels u-v are in presence of an isotropic field (bottom illustration). Each panel is normalized by the largest value. All parameters are the same as in Fig.~\ref{fig:2RL}.}
\label{fig:2RG}
\end{figure}
The simplest single plaquette provides a range of unexpected results, like the presence of extra symmetries with the magnetic field, invariance of the steady state w.r.t. magnetic field, etc., which help describe the complete nonequilibrium picture. We have also obtained the RDMs for the cubic system (both edge and face coupled, see Figs.~\ref{fig:2RL} and~\ref{fig:2RG}). The next example beyond a single plaquette, i.e., cube, increases the complexity drastically. The edge coupled ladder cube (cube without links between the plaquettes in the $z$ direction) possesses $6$ steady states for $\vec{B}=(0,0,0)$, $5$ steady states for $\vec{B}=(0,0,B)$ (anisotropic field), and $1$ for $\vec{B}=(B,B,B)$ (isotropic field). The presence of an isotropic field destroys all open-system symmetries and gives a unique NESS, i.e., steady state that depends on the bath parameters. All the other states are pure states and are eigenstates of the system Hamiltonian. 

Next we study the face connected cube in Fig.~\ref{fig:2RG}. In this case the direction of the magnetic field plays a significant role with $5$ steady states being present without a magnetic field (panels a-j), $5$ steady states for an anisotropic field $\vec{B}=(0,0,B)$ (panels k-t) and only $1$ state for an isotropic field $\vec{B}=(B,B,B)$ (panels u-v). In each case, we have only one NESS and the rest of the states are pure states. Here even though the number of steady states without and with anisotropic field are the same the states possess very different symmetries as evident by comparing first two rows of Fig.~\ref{fig:2RG} with the third and fourth row. This clearly shows that the direction of the magnetic field can play a very significant role on the structure of the steady states even though it may not influence their number.

In general, under the presence of the isotropic magnetic field the symmetries subspace merge and the nonequilibrium steady state changes accordingly. On the other hand, the magnitude of the magnetic field does not play a relevant role in the form of the NESS. To make this claim concrete, we compare the NESSs with and without a magnetic field using the fidelity, defined as 
\begin{equation}
\mathcal{F}\left( \rho_1,\rho_2 \right) = \Tr{ \sqrt{ \sqrt{\rho_1} \rho_2 \sqrt{\rho_1} } }.
\end{equation}
\begin{figure}[t!]
\centering
	\includegraphics[width=0.6\columnwidth]{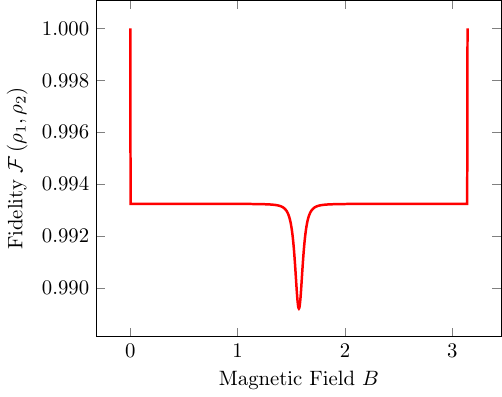}
\caption{Fidelity between the NESSs of the single plaquette (square) system with and without magnetic field (edge coupling). All parameters are the same as the ones used in Fig.~\ref{fig:analy}.}
\label{fig:fid-1R}
\end{figure}

In Fig.~\ref{fig:fid-1R} we display the fidelity of the square system with edge coupling to the baths. When the magnitude of the magnetic field is $B=n\pi$, with $n$ integer, the fidelity is maximum as the RDMs are the same. For a finite but not zero value of $B$ there is a discontinuous change in the value of the fidelity due to the symmetry breaking of the Liouvillian. Even after breaking the symmetry the value of the fidelity is always close to one, as the RDMs are similar. This shows that the NESS is robust under the presence of the magnetic field. The main difference appears when $B=\pi/2$, but even in this case the fidelity is almost $0.99$. In case of the single plaquette face connected system the fidelity always remains at unity, indicating that the NESS is invariant under the application of a magnetic field.

\begin{figure}[t!]
\centering
	\includegraphics[width=0.9\columnwidth]{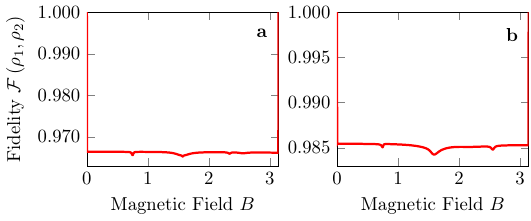}
\caption{Fidelity between the NESSs of the 2-plaquette (cube) system without magnetic field and with isotropic magnetic field of magnitude $B$. Fidelity for the edge coupled case is shown in panel a and face coupled case in panel b. All parameters are the same as the ones used in Fig.~\ref{fig:2RL}.}
\label{fig:fid-2R}
\end{figure}

For the cube scenario the fidelity shows a slightly more complicated behaviour as shown in Fig.~\ref{fig:fid-2R}. In the edge-coupling case (panel a) the difference in fidelity between having no field and a non-zero field is bigger than in the single plaquette case, but it is still very small $\approx 1$ (similar observations holds true for face coupled case, panel b). Besides, the fidelity presents several local minima, all of them only slightly lower than the flat value. Thus, the NESS is changed because once there is a magnetic field it does not have well defined symmetries, but it is still very similar to the non-magnetic field case implying that the \emph{magnitude} of the field itself plays a minimal role.

For completeness, we have also analyzed a {\it stacked} case, i.e. a cubic network with no connections in the $z$-direction (Fig.~\ref{fig:2RS}). In this case we found $8$ different steady states (panels a-p), where only one is a NESS (panels a-b), $5$ others are pure-states and there are also $2$ non-physical RDMs (panels g-h), meaning that they have zero trace. The existence of these non-physical solutions to a degenerated Lindblad master equation was hypothesised in Ref.~\cite{BucaNJP12} but no examples had been found until now. The bottom two rows correspond the presence of an isotropic or anisotropic field $\vec{B}=(0,0,B)$ or $\vec{B}=(B,B,B)$. Since the vertical links are absent in the stacked case the $x$- and $y$-direction magnetic fields play no role, since these only affect the vertical couplings. In this case again we only have one NESS (panels q-r) whereas all the others are pure states. The pure states show a pattern of symmetries in the simple cubic lattice far more complex than the single plaquette (square) lattice shown in Fig.~\ref{fig:analy}.

To summarize this section, the main idea is to study simple models that are analytically tractable. We chose to present the RDMs visually instead of presenting the complicated analytic solutions obtained using the method described in~\ref{sec:degenerated}. In these simple cases, we evaluated the initial condition independent steady state and found that breaking closed-system symmetries affects the open system properties. Moreover, the RDMs are affected mainly by the direction of the magnetic field and not by its magnitude (see Figs.~\ref{fig:fid-1R},~\ref{fig:fid-2R}: Fidelity deviations from the flat line when a magnetic field is applied). The face coupled system elucidated the role of the system-bath coupling operator and showed that the presence of magnetic field can give rise to non-topological symmetry. Here, even though the presence of magnetic field breaks open-system symmetries it does not affect all RDMs equally. In this case, we observed that the NESS remains invariant even though the other pure-state RDMs were affected. Our simple model also allowed us to find non-physical states which could be identified in the stacked two-plaquette structure with edge coupling to the baths. Thus, our simple model helps us to unravel some of the important consequences of open-system symmetries.

\section{Cubic Slab}
\label{sec:numerics}
Increasing in the network size greatly increases the complexity making it impossible to obtain analytic results beyond the two plaquette system. Hence, we address this issue numerically by calculating the number of steady states. It is worth noting that even obtaining the initial condition independent steady state \textit{density matrices} numerically is highly non-trivial task for a degenerated system since the numerical output can be any linear combination which could violate positivity (See~\ref{sec:degenerated}).

\begin{figure}[t]
\centering
	\includegraphics[width=1\columnwidth]{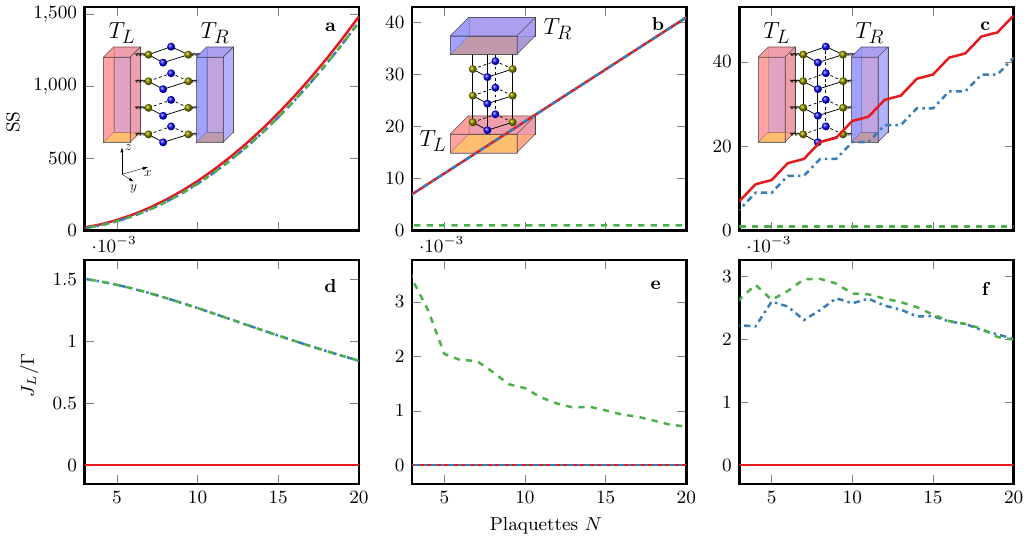}
	\caption{(Color online) Number of steady states (SS) (panels a-c) and particle currents ($J_L/\Gamma$) (panels d-f) for the simple cubic open systems illustrated in the insets of panels a-c. Red solid line corresponds to $\vec{B}=(0,0,0)$ (maximum symmetries), blue dash-dotted line is $\vec{B}=(0,0,B)$, and green dashed line is $\vec{B}=(B,B,B)$ (minimum symmetries). In panels a and d, blue dash-dotted and green dashed lines overlap. In panel b and e, red solid and blue dash-dotted lines overlap. The other parameters are the same as Fig.~\ref{fig:analy}.}
\label{fig:3D}
\end{figure}

Here, we consider three cases of a cubic system illustrated in Fig.~\ref{fig:3D}. The \emph{stacked} system consists of $N$ disconnected plaquettes with each plaquette connected to the reservoir via an edge coupling as shown in Fig.~\ref{fig:3D}a (inset). The ladder systems have inter-plaquette connections but a dichotomy exists depending on the system-reservoir coupling. The reservoir can either connect to the faces of the first and last plaquette of the ladder (inset panel b), i.e., \emph{face connected ladder}, or can connect via the edges (inset panel c), i.e., \emph{edge connected ladder}. In the absence of magnetic field the system can have weak and strong symmetries \cite{ManzanoAdvPhys18} that are analytically intractable. Hence to gain insight into the complex behaviour of the number of symmetry operators on the number of plaquettes in the system we plot the number of steady states (SS) in the top row of Fig.~\ref{fig:3D}.

In the edge connected stacked system, without the constraint of the inter-plaquette connections, the plaquettes are free to permute. This freedom leads to a faster than exponential growth (according to Sterling approximation) in the number of symmetries that is reflected in the fast growing number of steady states as seen in panel a (red solid line) for zero magnetic field $\vec{B}=(0,0,0)$. The intra-plaquette symmetry, i.e., flip along the diagonal of a plaquette as discussed previously, plays a rather insignificant role as compared to the symmetry generation due to inter-plaquette permutations. Hence, even when we break the intra-plaquette symmetry by applying an anistropic magnetic field along the $z$-direction $\vec{B}=(0,0,B)$ [blue dotted line in panel a] we still observe an exponential-like growth with slightly smaller number of steady states. Here due to the absence of intra-plaquette connections the isotropic magnetic field scenario is equivalent to the ansiotropic case (blue dotted and green dashed lines overlap). The presence of multiple steady states also affects the nonequilibrium observables including the steady-state particle currents ($J_L = \mathrm{Tr} \left[\left(A_{L1}^{\dagger}A_{L1}-A_{L2}^{\dagger}A_{L2}\right)\rho_{ss}\right]$) \cite{ManzanoPRB14} as shown in the bottom row of Fig.~\ref{fig:3D}. Since the direction of the magnetic field affects the symmetries, it also affects the currents in the system with the minimum current obtained when we have maximum number of steady states $\vec{B}=(0,0,0)$. This is simply because the initial condition is a complete dark state of the system $\rho(0)\propto \sum_{n=2,4,\cdots}|n\rangle\langle n|-|n\rangle\langle n+2|-|n+2\rangle\langle n|$ under intra-plaquette flip symmetry which is an eigenstate only when there is no applied magnetic field. In this case only the presence or absence of the magnetic field affects the transport creating a switch that can turn the current on or off.

The face connected ladder system shows a linear dependence between the number of symmetries and the number of plaquettes $N$. Since the system is placed in a dissipative gradient the inter-plaquette permutations are forbidden and do not contribute to multiple steady states. Hence, in this case the main contribution to the number of steady states comes from the intra-plaquette exchanges. The intra-plaquette symmetry that matters is the $180^\circ$ rotation, which is preserved when an anisotropic field $\vec{B}=(0,0,B)$ is applied to the system. Hence, as seen in panel b, the number of steady states remains invariant in the zero field (red solid line) to anisotropic field (blue dotted line) case. When an isotropic magnetic field is applied in this case all symmetries are broken and the system has only one unique steady state (green dashed line). In this case, the current obeys Fouriers law due to the nonlinear coupling with the reservoir even though the system is ballistic \cite{ManzanoNJP15}. As compared to the edge connected stacked system, where the mere presence of a magnetic field produced nonzero currents, here the direction of the magnetic field strongly affects the transport properties and only an isotropic field is able to produce finite currents.

The edge connected ladder system displays a highly complex behavior with stepwise \cite{note3} increase in the number of steady states as a function of the number of plaquettes Fig.~\ref{fig:3D}c. The stepwise behavior is present for an anisotropic field or in the absence of a field, but not in an isotropic field. Hence, the emergence of this behavior seems to be a result of inter-plaquette symmetries. The step-like behavior causes the currents to fluctuate, but as the number of plaquettes $N$ increase the anisotropic and isotropic magnetic field show the same behaviour. Overall, in the three systems in Fig.~\ref{fig:3D}, different symmetries are broken as a result of applying an anisotropic and isotropic magnetic field. This leads to different manipulation mechanisms of nonequilibrium currents with the presence (edge connected stacked) and direction (face connected ladder) of the magnetic field along with the number of plaquettes (edge connected ladder) playing important controls.

\section{Conclusions}
\label{sec:conclusion}
To summarize, this work proposes the control of nonequilibrium properties of a quantum cubic network using a magnetic field. Unlike traditional approaches such as the Aharonov-Bohm effect where the magnitude of the magnetic field is used to control transport \cite{EngelhardtPRB19, RaiJPCL11}, we provide a fresh perspective to utilize the \emph{direction} of the magnetic field. Specifically, this paper shows how a simple tight-binding lattice with \emph{engineered} couplings to leads could possess symmetries which can be broken systematically by the direction of the applied magnetic field. The symmetry breaking controls not only the steady states but also observables like the currents. Our results do not depend on the choice of parameters such as the magnitude of the magnetic field, coupling strength to the environment, or the specific values of the hopping and on-site potentials as long as the symmetries are maintained. Although this paper focuses on the transport in simple cubic networks, one can envision similar implementations in optics \cite{LeykamPRL18} or in new materials such as multi-layer graphene \cite{shalil:ssc12} that would extend our work to systems with hexagonal symmetries. We expect our work to trigger significant developments in understanding the connections between various lattice symmetries and open-system symmetries in order to manipulate quantum transport at the nanoscale. A promising research direction would be to utilize the presence of pure states to store energy and strategically release this energy using magnetic field controls to form quantum batteries \cite{Barra19}.

\section*{Acknowledgements}
This research was supported by the Institute for Basic Science in Korea (IBS-R024-Y2), DM acknowledges the Spanish Ministry and the Agencia Espa{\~n}ola de Investigaci{\'o}n (AEI) for financial support under grant FIS2017-84256-P (FEDER funds), and JC acknowledges support from NSF grant 1800301 and 1836913.

J.T. and D.M. contributed equally to this work.

\appendix
\section{Master Equation in the single-particle manifold}
\label{sec:single-exc}
In this section, we will show how the general many particle representation could be reduced to the single-particle picture. The system Hamiltonain as given by Eq.~(\ref{eq:1}) reads,
\begin{equation}
\label{eq:A1}
H=-\sum_{r} \varepsilon_r c^{\dagger}_r c_r -\sum_{r, d =1,2,3} t_d U^d(\hat{r})c^{\dagger}_{r+d}c_r + \mathrm{h.c.}.
\end{equation}
The tight-binding system Hamiltonian can easily be expressed in the single particle Fock basis as,
\begin{equation}
\label{eq:A2}
H^1=-\sum_{r} \varepsilon_r |r\rangle\langle r| -\sum_{r, d =1,2,3} t_d U^d(\hat{r}) |r+d \rangle\langle r |  + \mathrm{h.c.}.
\end{equation}
Above, the state $|r\rangle \equiv |0 0 \cdots 1_r 0 \cdots\rangle$ is a single particle Fock state with one particle being present at the $r$th site. The Lindblad operators connect the single-particle and vacuum state manifolds and are given by,
\begin{equation}
A_{\alpha 1} \equiv  A_{\alpha 2}^\dagger =\sum_{n} c_n^\dagger \Pi_{i\neq n} c_ic_i^{\dagger},
\end{equation}
with the sum over $n$ being restricted to $n=1,5,9,\cdots$ ($n=1,\cdots,4$) for $\alpha = L$ and $n=3,7,11,\cdots$ ($n=N-3,\cdots,N$) for $\alpha = R$ for edge (face) connected baths. In the Fock basis representation the Lindblad operator $A_{\alpha 1}^{(SE)} = |n\rangle \langle 0 |$ with $\ket{0}$ representing the ground state. The form of the Lindbladians does not increase the system Hilbert space exponentially but rather only adds the vacuum state to the single-particle state space. In other words, the dimension of the system is $N+1$, $N$ being the number of sites, instead of single-particle dimension $N$. Thus, we take our invariant basis set ($N+1$ states) comprising of the single-particle basis and the vacuum state to construct a single excitation Hamiltonian $H^{(SE)}= H^1 + \varepsilon_0 |0\rangle\langle 0|$ with $\varepsilon_0$ being the vacuum state energy. The Hamiltonian ensures that there is \emph{at most} one excitation/particle in the system. In realistic systems, this could be a result of being at low temperatures wherein the probability of finding more than one excitation is very low. Therefore, the quantum master equation governing the single excitation reduced density matrix reads
\begin{equation}
\frac{d\rho^\textrm{(SE)}}{dt} = - i [H^\textrm{(SE)},\rho^\textrm{(SE)}] + \sum_{k=1,2 \atop \alpha={ L, R}} \Gamma_{\alpha k} \mathcal{D}[A^\textrm{(SE)}_{\alpha k}] \rho^\textrm{(SE)}.
\end{equation}
Since we are interested in single particle currents and symmetries of the density matrices, dealing with the single excitation reduced density matrix reduces the computation drastically allowing us to deal with large number of sites.

\section{Density matrices of a degenerated Liouvillian}
\label{sec:degenerated}
In the simple cubic structures we are tacking we have a quantum Liouvillian $\mathcal{L}$ with several zero eigenvalues. These eigenvalues are a consequence of the existence of one or several non-trivial symmetry operators, $\pi_i$, such that they commute with all the generators of the dynamics in the form $[\pi_i,H]=[\pi_i,A_{\alpha,k}]=0 \quad \forall (\alpha,k)$ (strong symmetry) or because of the existence of one or several superoperators, $\Pi_i$,  that commutes with the full Liouvillian operator (weak symmetry) \cite{BucaNJP12}. Finding the steady states of such Liouvillians is highly non-trivial, as by direct diagonalisation we can obtain eigenvectors of the Liouvillian that do not fulfill the requirement for being density matrices (semi-definite positivity). Only if we know a complete set of operators, i.e., the set $\{\pi_i,H,A_{\alpha,k}\}$ of all jump and symmetry operators that generate the entire algebra of the operators Hilbert space, we can calculate all the density matrices by block diagonalising the Liouvillian matrix.

In the worst-case scenario we have a Liouvillian with $N$ zero eigenvalues, but no information about the symmetries. This Liouvillian could be reduced to a smaller one if some but not all of the symmetries are known. As the degeneracy comes from the symmetry we know that there are $N$ Hermitian matrices $\rho_i$, s.t. $\mathcal{L} \rho_i=0$ (which we will refer as zero-eigenmatrices), belonging to different symmetry subspaces. It is easy to prove that these matrices should be orthogonal in the sense that $\Tr{\rho_i \,\rho_j}=0$ if  $i\ne j$. Most likely, these matrices would have non-zero trace and they may be semi-definite positive, but there is also the possibility of having matrices with zero trace if they belong to a subspace with mixed symmetries (see Refs \cite{BucaNJP12, ManzanoAdvPhys18} for discussion). We will not consider the latter case below, as it is untypical, but the method proposed here can be trivially extended to a case with zero-trace zero-eigenmatrices. 
By direct diagonalisation of the Liouvillian matrix we can obtain $N$ eigenmatrices, $\tilde{\rho}_i$, s.t. $\mathcal{L} \tilde{\rho}_i=0$. These matrices need not be Hermitian, semi-definite positive or have unit trace as they can be just a combination of the real density matrices, i.e., they do not belong to a specific symmetry subspace. Furthermore, they do not necessarily fulfill the orthogonality condition. The problem thus can be summarised as follow: We have a set of $N$ zero-eigenmatrices $\{\tilde{\rho}_i\}$ that are a linear combination of $N$ density matrices $\rho_i$, how can we recover the set $\{\rho_i\}$ from $\{\tilde{\rho}_i\}$?

First, we need the matrices to be Hermitian. This can be done by generating new matrices $\rho^H_i= \tilde{\rho}_i + \tilde{\rho}_i^\dagger$. Besides, we need the matrices to form an orthogonal set. We then apply a Gramm-Schmidt type algorithm in the following form;
\begin{eqnarray}
 \rho_1^o &=& \rho_1^H. \nonumber\\
 \rho_2^o &=& \rho_2^H- \frac{\Tr{\rho_1^o \; \rho_2^H}}{\Tr{\rho_1^o \;\rho_1^o }} \rho_1^o \nonumber\\
 \rho_3^o &=& \rho_3^H- \frac{\Tr{\rho_1^o \; \rho_3^H}}{\Tr{\rho_1^o \;\rho_1^o }} \rho_1^o - \frac{\Tr{\rho_2^o \; \rho_3^H}}{\Tr{\rho_2^o \;\rho_2^o }} \rho_2^o \nonumber\\
& & \vdots  \nonumber\\
 \rho_N^o &=& \rho_N^H - \sum_{j=1}^{N-1}  \frac{\Tr{\rho_j^o \; \rho_N^H}}{\Tr{\rho_j^o \;\rho_j^o }} \rho_j^o. 
\end{eqnarray}
The set $\{\rho_i^o\}$ is now a set of orthogonal zero-eigenmatrices of the Liouvillian. The only problem remaining is that this set may contain non-semi-definite positive matrices, meaning that the eigenvalues of some of the elements of $\{\rho_i^o\}$ may be negative. To transform them in a set of positive zero-eigenmatrices and preserving the orthogonality of the system we may apply a rotation in the $N$ real space. We define a column vector of the zero-eigenmatrices $\vec{\rho}^{\,0}:=\left( \rho_1^o, \,\rho_2^o,\cdots,\,\rho_N^o \right)^T$ and the rotation is given by a unitary matrix $U_N(\theta_1,\theta_2,\cdots,\theta_k)$ with $\{\theta_1,\theta_2,\cdots,\theta_k\}$ being a set of Euler angles with $k=\frac{N^2-N}{2}$. For a fixed set of angles we can obtain a new set of rotated eigenmatrices as $\vec{\rho}(\theta_1,\theta_2,\cdots,\theta_k)=U_N(\theta_1,\theta_2,\cdots,\theta_k) \vec{\rho}^{\,0}$.

To fix the parameters to find the rotation that makes all the density matrices positive we define a function 
\begin{equation}
F(\theta_1,\theta_2,\cdots,\theta_k) = \sum_{i=1}^{N} \sum_{j=1}^d  \lambda_{i,j}(\theta_1,\theta_2,\cdots,\theta_k) - \left| \lambda_{i,j}(\theta_1,\theta_2,\cdots,\theta_k) \right| ,
\end{equation}
with $d$ being the dimension of the density matrices space and $ \lambda_{i,j}(\theta_1,\theta_2,\cdots,\theta_k)$ being the $j$th eigenvalue of the $i$th matrix from the vector $\vec{\rho}(\theta_1,\theta_2,\cdots,\theta_k)$. It is clear that when the vector $\vec{\rho}(\theta_1,\theta_2,\cdots,\theta_k)$ contains only positive matrices if the corresponding function $F(\theta_1,\theta_2,\cdots,\theta_k)$ is equal to zero. Therefore, we can solve the equation $F(\theta_1^*,\theta_2^*,\cdots,\theta_k^*)=0$, or maximize the function $F$, in order to find the angles that transform our set into a set of positive matrices. 
\begin{equation}
\vec{\rho}^{\,P} = U(\theta_1^*,\theta_2^*,\cdots,\theta_k^*) \vec{\rho}^{\,0}.
\end{equation}
Finally, we normalise each matrix in the vector $\vec{\rho}^{\,P}$ by doing $\rho_i=\rho^{P}_i /\Tr{ \rho^{P}_i}$ obtaining the desired set of density matrices. 

By the use of this method we can calculate the steady-states of any degenerated Liouvillian without knowing the symmetry operators. The main difficulty of this method is to perform a minimization over $\frac{N^2-N}{2}$ angles, being $N$ the degeneracy number, and the calculation of the function $F$ that requires the diagonalization of $N$ matrices. For big, highly degenerated systems the computational problem is very hard  and the use of efficient numerical algorithms for minimisation and diagonalisation  is required. For small systems with a low degeneration it is possible to find the steady-states, as we did with all the plots of this paper. Finally, if some but not all of the symmetry operators are known it is possible to reduce the complexity of the problem by applying the orthonormalisation algorithm only to the symmetry eigenspaces.


\section*{References}
\begin{thebibliography}{100}
\expandafter\ifx\csname url\endcsname\relax
  \def\url#1{\texttt{#1}}\fi
\expandafter\ifx\csname urlprefix\endcsname\relax\def\urlprefix{URL }\fi
\providecommand{\bibinfo}[2]{#2}
\providecommand{\eprint}[2][]{\url{#2}}

\bibitem{BucaNJP12}
Bu\v{c}a B and Prosen T 2012 New J. Phys. \textbf{14} 073007

\bibitem{ManzanoPRB14}
Manzano D and Hurtado P I 2014 Phys. Rev. B \textbf{90} 125138

\bibitem{ManzanoAdvPhys18}
Manzano D and Hurtado P I 2018 Adv. Phys. \textbf{6} 1

\bibitem{ThingnaSciRep16}
Thingna J, Manzano D and Cao J 2016 Sci. Rep. \textbf{6} 28027

\bibitem{MacieszczakPRL16}
Macieszczak K, Gu\c{t}\u{a} M, Lesanovsky I and Garrahan J P 2016 Phys. Rev. Lett. \textbf{116} 240404

\bibitem{AlbertPRA14}
Albert V V and Jiang L 2014 Phys. Rev. A \textbf{89} 022118

\bibitem{Ashcroft76}
Ashcroft N W and Mermin N D 1976 \emph{Solid State Physics} (Harcout, Orlando)

\bibitem{SchneiderNatPhys12}
Schneider U, Hackerm\"{u}ller L, Ronzheimer J P, Will S, Braun S, Best T, Bloch I, Demler E, Mandt S, Rasch D and Rosch A 2012 Nat. Phys. \textbf{8} 213

\bibitem{HildPRL14}
Hild S, Fukuhara T, Schau P, Zeiher J, Knap M, Demler E, Bloch I and Gross C 2014 Phys. Rev. Lett. \textbf{113} 147205

\bibitem{KumphNJP11}
Kumph M, Brownnutt M and Blatt R 2011 New J. Physics \textbf{13} 073043

\bibitem{BrittonNat12}
Britton J W, Sawyer B C, Keith A C, Wang C C J, Freericks J K, Uys H, Biercuk M J and Bollinger J J 2012 Nature \textbf{484} 489

\bibitem{WilsonNat14}
Wilson A C, Colombe Y, Brown K R, Knill E, Leibfried D and Wineland D J 2014 Nature \textbf{512} 57

%
%

\bibitem{ZnidaricPRL13}
\v{Z}nidari\v{c} M 2013 Phys. Rev. Lett. \textbf{110} 070602

\bibitem{ManzanoNJP15}
Manzano D, Chuang C and Cao J 2015 New J. Physics \textbf{18} 043044

%
%

\bibitem{GoldmanRPP14}
Goldman N, Juzeli\"{u}nas G, Ohberg P, Spielman I B 2014 Rep. Prog. Phys. \textbf{77} 126401

\bibitem{note1}
The magnetic phases are unphysical quantities but the flux $\Phi= 2\pi B$ that pierces each plaquette (in accordance with Stokes theorem) is physical.

\bibitem{HasegawaJPSJ90}
Hasegawa Y 1990 J. Phys. Soc. Jap. \textbf{59} 4384

\bibitem{HasegawaPhysicaC91}
Hasegawa Y 1991 Physica C \textbf{185} 1541

\bibitem{LinPRB96}
Lin Y-L and Nori F 1996 Phys. Rev. B \textbf{53} 13374

\bibitem{BurrelloJPA17}
Burrello M, Fulga I C, Lepori L and Trombettoni A 2017 J. Phys. A: Math. Theor. \textbf{50} 455301

\bibitem{Burrelloreview17}
Burrello M, Lepori L, Paganelli S and Trombettoni A 2017 \emph{Abelian Gauge Potentials on Cubic Lattices} (Advances in Quantum Mechanics: Contemporary Trends and Open Problems, G. Dell'Antonio and A. Michelangeli eds., Springer-INdAM series)

\bibitem{GeislerPRL04}
Geisler M C, Smet J H, Umansky V, von Klitzing K, Naundorf B, Ketzmerick R and Schweizer H 2004 Phys. Rev. Lett. \textbf{92} 256801

\bibitem{MelintePRL04}
Melinte S, Berciu M, Zhou C, Tutuc E, Papadakis S J, Harrison C, De Poortere E P, Wu M, Chaikin P M, Shayegan M, Bhatt R N and Register R A 2004 Phys. Rev. Lett. \textbf{92} 036802

\bibitem{DeanNat13}
Dean C R, Wang L, Maher P, Forsythe C, Ghahari F, Gao Y, Katoch J, Ishigami M, Moon P, Koshino M, Taniguchi T, Watanabe K, Shepard K L, Hone J and Kim P 2013 Nature \textbf{497} 598

\bibitem{PonomarenkoNat13}
Ponomarenko L A, Gorbachev R V, Yu G L, Elias D C, Jalil R, Patel A A, Mishchenko A, Mayorov A S, Woods C R, Wallbank J R, Mucha-Kruczynski M, Piot B A, Potemski M, Grigorieva I V, Novoselov K S, Guinea F, Falko V I and Geim A K 2013 Nature \textbf{497} 594

\bibitem{AidelsburgerPRL13}
Aidelsburger M, Atala M, Lohse M, Barreiro J T, Paredes B and Bloch I 2013 Phys. Rev. Lett. \textbf{111} 185301

\bibitem{MiyakePRL13}
Miyake H, Siviloglou G A, Kennedy C J, Burton W C and Ketterle W 2013 Phys. Rev. Lett. \textbf{111} 185302

\bibitem{DiehlNatPhys08}
Diehl S, Micheli A, Kantian A, Kraus B, B\"{u}chler H P and Zoller P 2008 Nat. Phys. \textbf{4} 878

\bibitem{MullerAAMOP12}
M\"{u}ller M, Diehl S, Pupillo G and Zoller P 2012 Adv. At. Mol. Opt. Phys. \textbf{61} 1

\bibitem{Alicki2007}
Alicki K and Lendi R 2007 \emph{Quantum Dynamical Semigroups and Applications} (Springer-Verlag, Berlin/Heidelberg)

\bibitem{Breuer2007}
Breuer H and Petruccione F 2007 \emph{The Theory of Open Quantum Systems} (OUP, Oxford)

\bibitem{Gorini1976}
Gorini V, Kossakowski A and Sudarshan E C G 1976 J. Math. Phys. \textbf{17} 821

\bibitem{Lindblad1976}
Lindblad G 1976 Commun. Math. Phys. \textbf{48} 119

\bibitem{EngelhardtPRB19}
Engelhardt G and Cao J 2019 Phys. Rev. B \textbf{99} 075436

\bibitem{MaimaitiPRB17}
Maimaiti W, Andreanov A, Park H C, Gendelman O and Flach S 2017 Phys. Rev. B \textbf{95} 115135

\bibitem{note3}
In the plateau region the number of steady states increases by 1 with increasing plaquettes.

\bibitem{RaiJPCL11}
Rai D, Hod O and Nitzan A 2011 J. Phys. Chem. Lett. \textbf{2} 2118

\bibitem{LeykamPRL18}
Leykam D, Mittal S, Hafezi M and Chong Y D 2018 Phys. Rev. Lett. \textbf{121} 023901

\bibitem{shalil:ssc12}
Shahil K M F and Balandin A A 2012 Solid State Communications \textbf{152} 1331

\bibitem{Barra19}
Barra F 2019 Phys. Rev. Lett. \textbf{122} 210601

\bibitem{Evans}
 D.E. Evans 1977 Commun. Math. Phys. \textbf{54} 293
 
\end{thebibliography}
\end{document}